\documentclass[prd,twocolumn,amsfonts,amssymb,nofootinbib]{revtex4}

\usepackage{graphicx}

\usepackage{bm}

\newcommand{\beq}{\begin{equation}}
\newcommand{\eeq}{\end{equation}}
\newcommand{\beqs}{\begin{eqnarray}}\newcommand{\eeqs}{\end{eqnarray}}
\newcommand{\lsim}{\mathrel{\raisebox{-
.6ex}{$\stackrel{\textstyle<}{\sim}$}}}
\newcommand{\gsim}{\mathrel{\raisebox{-
.6ex}{$\stackrel{\textstyle>}{\sim}$}}}

\begin{document}

\title{Improved Upper Limits on Baryon-Number Violating Dinucleon Decays to
  Dileptons}

\author{Sudhakantha Girmohanta and Robert Shrock}

\affiliation{ \ C. N. Yang Institute for Theoretical Physics and 
Department of Physics and Astronomy, \\
Stony Brook University, Stony Brook, NY 11794, USA }

\begin{abstract}

  We consider effects of $n-\bar n$ oscillations and resultant matter
  instability due to dinucleons decays. We point out that
  existing upper bounds on the rates for the dinucleon decays $nn \to 2\pi^0$,
  $nn \to \pi^+\pi^-$, and $np \to \pi^+\pi^0$ imply upper bounds on the rates
  for dinucleon decays to dileptons $nn \to e^+ e^-$, $nn \to \mu^+\mu^-$, $nn
  \to \nu_\ell \bar\nu_\ell$, and $np \to \ell^+ \nu_\ell$, where $\ell=e, \
  \mu, \ \tau$.  We present estimates for these upper bounds.  Our bounds 
  are substantially stronger than corresponding limits from direct searches.

\end{abstract}

\maketitle

% =======================================================================

% section 1
\section{Introduction}
\label{intro_section}

The violation of baryon number, $B$, is expected to occur in nature, because
this is one of the necessary conditions for generating the observed baryon
asymmetry in the universe \cite{sakharov}.  Baryon number violation (BNV) is,
indeed, predicted in many ultraviolet extensions of the Standard Model (SM),
such as grand unified theories. A number of dedicated experiments have
been carried out since the early 1980s to search for proton decay (and the
decay of neutrons bound in nuclei). These experiments have obtained null
results and have set resultant stringent upper limits for the rates of such
$\Delta B=-1$ baryon-number-violating nucleon decays.

A different type of baryon number violation has also received attention, namely
$n - \bar n$ oscillations, which have $|\Delta B|=2$
\cite{kuzmin}-\cite{sk_nnbar}. It was observed early on that $n-\bar n$
oscillations might provide the source of baryon number violation necessary for
baryogenesis \cite{kuzmin}.  The same operators that mediate $n-\bar n$
transitions also lead to matter instability via the dinucleon decays from $nn$
and $np$ initial states to respective multipion final states. Let us denote the
low-energy effective Hamiltonian responsible for $n-\bar n$ oscillations as
${\cal H}^{(n \bar n)}_{eff}$. We will assume a minimal framework in which
${\cal H}^{(n \bar n)}_{eff}$ incorporates all of the physics beyond the
Standard Model relevant for $n-\bar n$ oscillations.  Rates for these dinucleon
decays in matter are calculated by taking into account that in the presence of
a nonzero transition amplitude $\langle \bar n |{\cal H}^{(n \bar
  n)}_{eff}|n\rangle$, the physical state $|n \rangle_{phys.}$ contains a small
but nonzero $|\bar n\rangle$ component. This leads to a nonzero amplitude for
annhilation of the $|\bar n\rangle$ component with a neighboring neutron or
proton in a nucleus.

The operators in the low-energy effective Hamiltonian for proton decay are
four-fermion operators with Maxwellian mass dimension 6 and hence coefficients
of mass dimension $-2$, whereas the operators in ${\cal H}^{(n \bar
  n)}_{eff}$ are six-quark operators, with coefficients of dimension $-5$. 
Consequently, if one were to assume that there is a single high
mass scale $M_{BNV}$ characterizing the physics responsible for baryon number
violation, proton decay would be much more important as a manifestation of
baryon number violation than $n-\bar n$ oscillations and the corresponding
dinucleon decays.  However, such an assumption of a single BNV mass scale may
well be overly simplistic.  Ref. \cite{nnb02} presented an explicit example of
a theory in which proton decay is suppressed well beyond observable levels
while $n-\bar n$ oscillations occur at levels comparable to existing
experimental limits.  In such a model, it is the $n-\bar n$ oscillations and
the corresponding $nn$ and $np$ dinucleon decays to multi-meson final states
that are the main manifestations of baryon number violation, rather than
individual proton and bound neutron decays. Further examples of models 
with baryon number violation but no proton decay were given in the later work 
\cite{wise}. 

Here we point out that existing upper bounds on the rates for the hadronic
dinucleon decays $nn \to 2\pi^0$, $nn \to \pi^+\pi^-$, and 
$np \to \pi^+\pi^0$ imply upper bounds on
the rates for the dinucleon to dilepton decays $nn \to e^+ e^-$, $nn \to
\mu^+\mu^-$, $nn \to \nu_\ell \bar\nu_\ell$, and $np \to \ell^+ \nu_\ell$,
where $\ell=e, \ \mu, \ \tau$.  We present estimates for these upper bounds.
Our upper bounds are considerably stronger than direct limits on the rates for
these decays.

% ====================================================================

\section{$n - \bar n$ Oscillations and Dinucleon Decays to Hadronic Final
  States} 
\label{nnbar_section}

We recall some basic results on $n-\bar n$ oscillations that are needed for our
analysis (for further details, see, e.g., \cite{nnbar_white}). Let us consider
a general theory in which there is baryon-number violating physics beyond the
Standard Model (BSM) that leads to $n-\bar n$ transitions and let us denote the
corresponding transition amplitude as
\beq
\delta m = \langle \bar n | {\cal H}^{(n\bar n)}_{eff} | n \rangle \ . 
\label{deltam}
\eeq
In (field-free) vacuum, one is thus led
to diagonalize the matrix of the Hamiltonian in the basis $(|n\rangle, |\bar n
\rangle)$,
\beq
\left(\begin{array}{cc}
m_n - i\lambda_n/2 & \delta m \\
\delta m         & m_n - i\lambda_n/2 \end{array}\right) \ , 
\label{nnb_matrix}
\eeq
where $\lambda_n = \tau_n^{-1}$ is the decay rate of the free neutron and the
equality $m_{\bar n} = m_n$ follows from CPT invariance.  The eigenstates of
this matrix are $|n_\pm \rangle =( |n \rangle \pm |\bar n \rangle )/\sqrt{2}$,
with mass eigenvalues $m_{\pm} = (m_n \pm \delta m) - i\lambda_n/2$.  Hence, if
one starts with a pure $|n\rangle$ state at $t=0$, then there is a finite
probability for it to be an $|\bar n\rangle$ at $t \ne 0$ given by
\beq
P(n(t)=\bar n) = |\langle \bar n|n(t) \rangle|^2 = [\sin^2(t/\tau_{n \bar n})]
e^{-\lambda_n t} \ , 
\label{nbprob}
\eeq
where $\tau_{n \bar n} = 1/|\delta m|$. The current limit on $\tau_{n \bar n}$
from an experiment with a neutron beam from a nuclear reactor at the Institut
Laue-Langevin (ILL) in Grenoble is $\tau_{n \bar n} \ge 0.86 \times 10^8$ sec, 
i.e., $|\delta m| = 1/\tau_{n \bar n} < 0.77 \times 10^{-29}$ MeV \cite{ill}.
(This and other limits discussed here are at the 90 \% confidence level.) 

For a neutron bound in a nucleus, the Hamiltonian matrix 
becomes 
\beq
\left(\begin{array}{cc}
m_{n,eff.}  & \delta m \\
\delta m         & m_{\bar n, eff.} \end{array}\right)
\label{hmat}
\eeq
with $m_{n,eff} = m_n + V_n$ and $m_{\bar n, eff.} = m_n + V_{\bar n}$, where
the nuclear potential $V_n$ is real, $V_n = V_{nR}$, but $V_{\bar n}$ has an
imaginary part: $V_{\bar n} = V_{\bar n R} - i V_{\bar n I}$.  In the presence
of the $n-\bar n$ mixing, the resultant physical eigenstate for the neutron
state in matter has a small component of $|\bar n\rangle$, i.e.,
\beq
|n\rangle_{\rm phys.}  = \cos\theta_{n\bar n} |n\rangle + 
                        \sin\theta_{n \bar n} |\bar n\rangle \ . 
\label{nphys}
\eeq
where $\tan(2\theta_{n \bar n}) = 2\delta m/|m_{n,eff}-m_{\bar n, eff}|$. 
In contrast to the situation in field-free vacuum, where
$\theta=\pi/4$ and the mixing is maximal, in matter, because the diagonal
elements of the Hamiltonian matrix are different, $|\theta| << 1$. However,
this is more than compensated for by the large number of nucleons in a proton
decay experiment such as SuperKamiokande (SK). The nonzero $|\bar n\rangle$
component in $|n\rangle_{\rm phys.}$ leads to annihilation with an adjacent
neutron or proton, and hence to the decays to zero-baryon, multi-meson final
states consisting dominantly of several pions, $nn \to {\rm pions}$
and $np \to {\rm pions}$. The rate characterizing matter instability (m.i.) due
to these dinucleon decays is
\beq
\Gamma_{m.i.} \equiv \frac{1}{\tau_{m.i.}} \simeq  
\frac{2(\delta m)^2 |V_{\bar n I}|}
{(V_{n R} - V_{\bar n R})^2 + V_{\bar n I}^2} \ . 
\eeq
Hence, $\tau_{m.i.} \propto (\delta m)^{-2} = \tau_{n \bar n}^2$.  A common
convention is to introduce a multiplicative factor $R$ and to write
$\tau_{m.i.} = R \, \tau_{n \bar n}^2$. Here, $R \sim O(10^2)$ MeV, or
equivalently, $R \simeq 10^{23}$ sec$^{-1}$, dependent on the nucleus. Lower
limits on $\tau_{m.i.}$ that yield equivalent lower
bounds on $\tau_{n \bar n}$ in the $10^8$ sec. range have been obtained from
the Kamiokande \cite{takita86}, Soudan \cite{chung02}, SNO (Sudbury Neutrino
Observatory) \cite{aharmim17}, and SK \cite{sk_nnbar}
experiments.  The best current limit on matter instability (from SK) is
\cite{sk_nnbar},
\beq
\tau_{m.i.} > 1.9 \times 10^{32} \ {\rm yr} \ ,
\label{tau_mi}
\eeq
and hence, taking into account the uncertainty in the calculation of $R \simeq
0.52 \times 10^{23}$ sec$^{-1}$ for the ${}^{16}$O nuclei in water
\cite{friedman_gal,nnbar_white}, the SK experiment has inferred the limit
\cite{sk_nnbar}
\beq
\tau_{n \bar n} > 2.7 \times 10^8 \ {\rm sec}, \ \ i.e., \ \ 
|\delta m| < 2.4 \times 10^{-30} \ {\rm MeV}.
\label{tau_nnb_sk_limit}
\eeq
(From this and the value $|m_{n,eff}-m_{\bar n, eff}| \sim 10^2$ MeV, 
it follows that $|\theta_{n \bar n}| \lsim 10^{-31}$.) 

There have also been searches for dinucleon decays to specific final
states. Reflecting the dominance of the strong interactions over the
electroweak interactions, these decays lead mainly to hadronic final states.
From null searches for the decays 
${}^{56}{\rm Fe} \to {}^{54}{\rm Fe} + \pi^+\pi^-$ \cite{frejus91},
${}^{16}{\rm O} \to {}^{14}{\rm O} + 2\pi^0$ \cite{sk_dinucleon_to_pions}, and
${}^{16}{\rm O} \to {}^{14}{\rm N} + \pi^+\pi^0$ \cite{sk_dinucleon_to_pions},
experiments have set upper bounds on the rates
$\Gamma_i$, or equivalently, lower bounds on the partial lifetimes
$(\tau_i/B_i)_i \equiv \Gamma_i^{-1}$ for these decays, where $B_i$ 
denotes a branching ratio. The experiments use the notational convention of
referring to these as $nn \to \pi^+\pi^-$, $nn \to 2\pi^0$, and 
$np \to \pi^+\pi^0$. We will follow
this convention, but note that a conversion would be necessary to compute the
rate for an individual pair of neighboring nucleons to undergo these
decays. The limit from the Fr\'ejus experiment \cite{frejus91} is 
\beq
(\tau/B)_{nn \to \pi^+\pi^-} > 0.7 \times 10^{30} \ {\rm yr}, 
\label{tau_nn_to_pipm}
\eeq
and the limits from the SK experiment \cite{sk_dinucleon_to_pions} are 
\beq
(\tau/B)_{nn \to \pi^0\pi^0} > 4.04 \times 10^{32} \ {\rm yr} 
\label{tau_nn_to_pi0pi0}
\eeq
and
\beq
(\tau/B)_{np \to \pi^+\pi^0} > 1.70 \times 10^{32} \ {\rm yr} \ . 
\label{tau_np_to_pippi0}
\eeq
We use the two more stringent bounds  (\ref{tau_nn_to_pi0pi0}) and 
(\ref{tau_np_to_pippi0}) for our analysis. 

% ========================================================================

\section{Dinucleon Decays to Dilepton Final States}
\label{dinucleon_dlepton_section}

The same baryon-number-violating physics that leads to $n-\bar n$ oscillations
and hence also the dinucleon decays $nn \to {\rm pions}$ and $np \to {\rm
  pions}$ also leads to dinucleon decays to leptonic final states, in
particular, to dileptons:
\beq
nn \to \ell^+ \ell^-  \quad {\rm for} \ \ell=e, \ \mu
\label{nn_to_ellellbar}
\eeq
\beq
nn \to \nu_\ell \bar\nu_\ell \quad {\rm for} \ \nu_\ell= \nu_e, \ \nu_\mu,
\ \nu_\tau 
\label{nn_to_nunubar}
\eeq
and
\beq
np \to \ell^+ \nu_\ell  \quad {\rm for} \ \ell=e, \ \mu, \ \tau \ . 
\label{np_to_ellnubar}
\eeq
As is evident, these are $\Delta B = -2$, $\Delta L=0$ decays, where $L$
denotes total lepton number.  We will derive upper bounds on the rates for
these decays by relating them to hadronic dinucleon decays and using the upper
bounds on rates for the latter.  We utilize a minimal theoretical framework for
our analysis, namely to assume the BSM physics responsible for the $n-\bar
n$ oscillations, but then apply only Standard-Model physics to derive these
relations.  With this framework, we identify and estimate the leading
contributions to these dinucleon decays to dileptons.  These contributions
involve amplitudes each of which consists of a combination of two parts: 
(a) The basic BNV part, involving a four-fermion operator resulting from
physics operative at a mass scale $M_{BNV} >> v$, where $v=250$ GeV is the 
electroweak-symmetry-breaking (EWSB) scale, and a second part involving 
SM physics, with a virtual timelike photon, $Z$, or $W$. 

\begin{figure}
  \begin{center}
    \includegraphics[height=6cm]{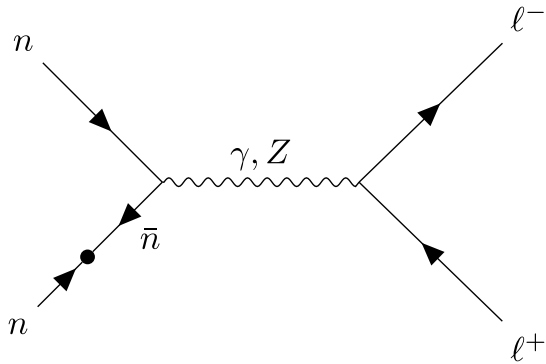}
  \end{center}
\caption{Feynman diagram for $nn \to \ell^+\ell^-$ with $\ell=e, \ \mu$.} 
\label{dinucleon_to_ellellbar_figure}
\end{figure}

We begin with the decay $nn \to \ell^+\ell^-$.  This decay can occur as
follows: the $|\bar n\rangle$ component in a $|n\rangle_{\rm phys.}$ neutron in
a nucleus leads to annihilation with a neighboring neutron to yield a virtual
photon in the $s$ channel, which then produces the final-state $\ell^+ \ell^-$
pair in (\ref{nn_to_ellellbar}). A much smaller contribution involves a diagram
with a virtual $Z$ in the $s$-channel. Equivalently, one can envision this as
being due to a transition in which an initial $n$ changes to a $\bar n$ with
transition matrix element (\ref{deltam}), and then the $\bar n$ annihilates
with the neighboring $n$ to produce the virtual photon or $Z$, as shown in
Fig. \ref{dinucleon_to_ellellbar_figure}.
Up to small corrections due to the bound state Fermi momenta of the nucleons,
the center-of-mass energy is $\sqrt{s}=m_n+m_p \equiv 2m_N$ in this transition,
and the $\ell^+$ and $\ell^-$ are emitted back-to-back, each with a total
energy in the lab frame equal to $m_N$. We denote the four-momentum of the
virtual photon or $Z$ as $q$ and the four-momenta of the $\ell^-$ and $\ell^+$
as $p_2$ and $p_1$, with $q=p_1+p_2$ and $q^2=s=(2m_N)^2$. Here and below, we
neglect small effects due to Fermi momenta.  

To leading order, the 
amplitude for $nn \to \ell^+\ell^-$ is the sum of the terms due to virtual 
($v$) photon and $Z$ exchange in the $s$-channel: 
\beq
A_{nn \to \ell^+ \ell^-} = 
A_{nn \to \ell^+ \ell^-;\gamma} + 
A_{nn \to \ell^+ \ell^-; Z} \ , 
\label{amp_nn_to_ellellbar}
\eeq
with
\beq
A_{nn \to \ell^+ \ell^-; \gamma} = 
(\delta m) \, e^2 \, \langle 0 | J_{em}^\lambda |n \bar n\rangle \, 
\frac{1}{q^2} \, [\bar u(p_2) \gamma_\lambda v(p_1)] 
\label{amp_nn_to_ellellbar_photon}
\eeq
and
\begin{widetext} 
\beq
A_{nn \to \ell^+ \ell^-; Z} = \sqrt{2} G_F \, (\delta m) \, 
\langle 0 | J_Z^\lambda |n \bar n\rangle \, 
 \Big [ \bar u(p_2)\gamma_\lambda [(1-4\sin^2\theta_W) - \gamma_5]v(p_1)
\Big ] \ ,
\label{amp_nn_to_ellellbar_z}
\eeq
\end{widetext}
where the $\delta m$ factor represents the initial $n \to \bar n$ transition
mediated by ${\cal H}^{(n \bar n)}_{eff}$; $J_{em}^\lambda$ and $J_Z^\lambda =
J_{3L}^\lambda - \sin^2\theta_W \, J_{em}^\lambda$ denote the electromagnetic
and neutral weak currents; and $e=\sqrt{4\pi \alpha_{em}}$, and $G_F$ denote
the electromagnetic and Fermi couplings.

We first consider the contribution from $A_{nn \to \ell^+ \ell^-; \gamma}$.
Since the annihilation occurs on a scale of order $\sim 1$ fm, a reasonable
approximation is to consider the initial $nn$ state by itself, independent of
the other nucleons in the nucleus. Let us denote the wavefunction of this state
as $|nn\rangle = \phi_I \, \phi_S \, \phi_L$, where $I$, $S$, and $L$ denote
the strong isospin, the spin, and the relative orbital angular momentum $L$ of
the $nn$ pair. (To maintain standard notation, we use the same symbol, $L$, for
orbital angular momentum and total lepton number; the context will always 
make clear which is meant.)  This wavefunction must be
antisymmetric under interchange of neutrons. The $|nn\rangle$ state has strong
isospin $I=1$, and the lowest-energy configuration has $L=0$, so the $\phi_I$
and $\phi_L$ wavefunctions for this configuration are both symmetric under
interchange of neutrons. Hence, $\phi_S$ is antisymmetric, corresponding to
spin $S=0$ and hence total angular momentum $J=0$ for the $nn$ pair.  Since
${\cal L}^{(n \bar n)}_{eff}$ is a Lorentz scalar, the $n-\bar n$ transition
matrix element $\langle \bar n | {\cal L}^{(n \bar n)}_{eff} | n\rangle$ does
not change the neutron spin, so the value of $S$ (as well as $L$) for the
resultant $n \bar n$ dinucleon is the same as for the initial $n n$
dinucleon. (This is obvious in Eq. (\ref{nphys}).)  The matrix element $\langle
0 | J_{em}^\lambda |n \bar n \rangle$ is related by crossing symmetry to the
matrix element $\langle n | J_{em}^\lambda | n \rangle$, which involves Dirac
and Pauli form factors $F^{(n)}_1(q^2)$ and $F^{(n)}_2(q^2)$. For the $J=0$
$nn$ state, the only four-momentum on which the matrix element $\langle 0 |
J_{em}^\lambda |n \bar n \rangle$ can depend is $q^\lambda$, so $\langle 0 |
J_{em}^\lambda |n \bar n\rangle \propto q^\lambda$.  But $q^\lambda \, [\bar
u(p_2) \gamma_\lambda v(p_1)] = 0$, so that this contribution to the amplitude
vanishes. Another contribution arises from an excited $|nn \rangle$ state with
$L=1$ and an antisymmetric $\phi_L$, so that $\phi_S$ is symmetric,
corresponding to $S=1$. Then the quantum mechanical addition of $L$ and $S$ to
yield a total angular momentum $\vec J = \vec L + \vec S$ can yield $J=0, 1$,
or 2. The $J=0$ state gives zero contribution, as before, so the amplitude
arises from the initial $nn$ states with nonzero $J$.  We denote the
probability of the $nn$ dinucleon to be in a state with $J \ne 0$ as $P_{nn,J
  \ne 0}$.  Given that $J \ne 0$ so that $A_{nn \to \ell^+\ell^-;\gamma} \ne
0$, it follows that in $|A_{nn \to \ell^+\ell^-;\gamma}|^2$, the $(1/s)^2$
factor from the photon propagator is cancelled by kinematic factors of order
$s^2$.

We next consider the contribution from $A_{nn \to \ell^+\ell^-; Z}$. The
square, $|A_{nn \to \ell^+\ell^-; Z}|^2$, is negligible because of suppression
by the factor $\sim (G_F s)^2 = 1.7 \times 10^{-9}$. The cross term ${\rm
  Re}\{A_{nn \to \ell^+\ell^-; \gamma} \, A^*_{nn \to \ell^+\ell^-; Z} \}$ is
also small because of the factor $\sim G_F s = 4.11 \times 10^{-5}$.  Thus,
although for the $J=0$ initial $nn$ state, the axial-vector part of $J_Z$ has a
nonzero contraction $q^\lambda [\bar u(p_2) \gamma_\lambda \gamma_5 v(p_1)] =
2m_\ell [\bar u(p_2)\gamma_5 v(p_1)]$, this contribution is suppressed both by
the smallness of $2m_\ell/\sqrt{s} = m_\ell/m_N$ and by the $G_F s$ factor in
the amplitude.

The two-body phase space factor for a decay of an initial state with 
mass $\sqrt{s}$ to final-state ($fs$) particles with masses $m_1$ and $m_2$ is 
\beq
R^{(fs)}2 = \frac{1}{8\pi} \, [ \lambda(1,m_1^2/s,m_2^2/s)]^{1/2} \ ,
\label{r2}
\eeq
where
\beq
\lambda(x,y,z)=x^2+y^2+z^2-2(xy+yz+zx) \ . 
\label{lam}
\eeq
Hence, for the relevant case
$m_1=m_2 \equiv m$, $R_2 = (8\pi)^{-1}\sqrt{1-4m^2/s}$. The square root is
equal to 0.9896, \ 1.0000, \ and 0.9937 for the respective decays
$nn \to 2\pi^0$, $nn \to e^+e^-$, and $nn \to \mu^+\mu^-$. 

We are thus led to the estimate
\beqs
\Gamma_{nn \to \ell^+\ell^-} & \sim & 
P_{nn,J \ne 0} \, e^4 \, 
\frac{R_2^{(\ell^+\ell^-)}}
     {R_2^{(2\pi^0      )}}    \, \Gamma_{nn \to 2\pi^0} \cr\cr
&\sim& P_{nn,J \ne 0} \, e^4 \, \Gamma_{nn \to 2\pi^0} \ , 
\label{gamma_nn_to_ellellbar}
\eeqs
where we have used the fact that $R_2^{(\ell^+\ell^-)}/R_2^{(2\pi^0)}$ is very
close to unity for both $\ell=e$ and $\ell=\mu$. 
Utilizing the lower limit on $(\tau/B)_{nn \to 2\pi^0}$ in
Eq. (\ref{tau_nn_to_pi0pi0}) together with the estimate
(\ref{gamma_nn_to_ellellbar}), we thus obtain the following estimates for
lower limits on the partial lifetimes for dinucleon to
dilepton decays per ${}^{16}$O nucleus:
\beqs
(\tau/B)_{nn \to \ell^+\ell^-} && \gsim 
(P_{nn,J \ne 0})^{-1} \, (5 \times 10^{34} \ {\rm yr})
 \cr\cr
&& \gsim 5 \times 10^{34} \ {\rm yr} \ {\rm for} \ \ell=e, \ \mu \ . 
\label{tau_limit_nn_to_ellellbar}
\eeqs
where the final inequality follows from the fact that $P_{nn,J \ne 0} <
1$. Even without inserting an estimated value for the suppression factor
due to $P_{nn,J \ne 0}$, our bound (\ref{tau_limit_nn_to_ellellbar}) is
stronger than the direct limits on these two decays, which are 
(from the SuperKamiokande experiment) \cite{sk_dinucleon_to_ellell}:
\beq
(\tau/B)_{nn \to e^+ e^-} > 4.2 \times 10^{33} \ {\rm yr}
\label{tau_limit_nn_to_ee_sk}
\eeq
and
\beq
(\tau/B)_{nn \to \mu^+ \mu^-} > 4.4 \times 10^{33} \ {\rm yr} \ . 
\label{tau_limit_nn_to_mumu_sk}
\eeq

\begin{figure}
  \begin{center}
    \includegraphics[height=6cm]{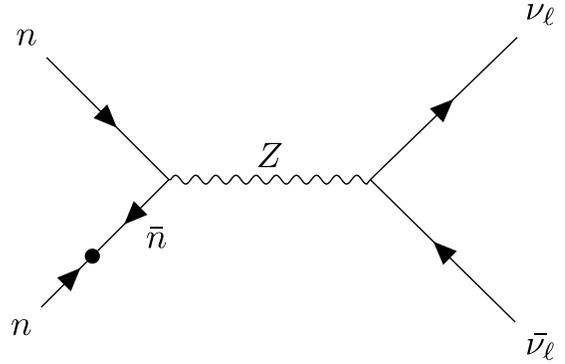}
  \end{center}
\caption{Feynman diagram for $nn \to \nu_\ell \bar\nu_\ell$, where 
$\nu_\ell = \nu_e, \ \nu_\mu, \ \nu_\tau$.} 
\label{dinucleon_to_nunubar_figure}
\end{figure}

We next consider the decay $nn \to \nu_\ell \bar\nu_\ell$, 
where $\nu_\ell = \nu_e, \ \nu_\mu$, or $\nu_\tau$.  This decay
arises from a process in
which the $|\bar n\rangle$ in $|n\rangle_{\rm phys.}$ annihilates with a
neighboring neutron to produce a virtual $Z$ boson in the $s$-channel, which
then yields the final-state $\nu_\ell \bar\nu_\ell$ pair, as shown in Fig. 
\ref{dinucleon_to_nunubar_figure}.  Here and below, we shall refer to this 
as a tree-level process, having integrated out any loops 
in a BSM model to obtain the local four-fermion operators in the low-energy 
effective Hamiltonian ${\cal H}^{(n \bar n)}_{eff}$.  (More precisely, it is a
tree-level process as regards SM fields.) 
One may again analyze the contributions of the $J=0$ and $J\ne 0$ initial $nn$
states. For the $J=0$ initial state, by the same argument as above, the vector
part of the neutral current gives a vanishing contribution, and the axial
vector part gives a negligibly small contribution to the amplitude proportional
to neutrino masses. Hence, the decay arises from the $J \ne 0$ initial
dineutron states.  We thus obtain the rough estimate 
\beq
\Gamma_{nn \to \nu_\ell \bar\nu_\ell} 
\sim P_{nn, J \ne 0} \, (G_F s)^2 \, \Gamma_{nn \to {\rm hadrons}} \ . 
\label{gamma_nn_to_nunubar}
\eeq
Combining this with the experimental limit (\ref{tau_nn_to_pi0pi0}), 
we obtain the rough lower bound, per ${}^{16}$O nucleus, 
\beqs
(\tau/B)_{nn \to \nu_\ell \bar\nu_\ell} && 
\gsim P_{nn,J \ne 0}^{-1} \, (2 \times 10^{41} \ {\rm yr}) \cr\cr
&& \gsim 2 \times 10^{41} \ {\rm yr} 
\quad {\rm for} \ \nu_\ell= \nu_e, \ \nu_\mu, \ \nu_\tau \ . 
\cr\cr
&&
\label{tau_limit_nn_to_nunubar}
\eeqs
For comparison, there is a bound from a direct search by the KamLAND
experiment\footnote{ The KamLAND bound was
  obtained via a search for the decays of the resultant ${}^{10}$C nucleus
  \cite{kamland_nn_to_nunubar}. Although our bound applies to an ${}^{16}$O
  nucleus rather than ${}^{12}$C nucleus, one does not expect the rates to
  differ very much between these nuclei with almost equal numbers of
  nucleons. A weaker bound, $(\tau/B)_{nn \to {\rm inv.}} > 1.3 \times 10^{28}$
  yr. per ${}^{16}$O nucleus has been obtained by the SNO+ experiment
\cite{sno_invisible}.}, namely \cite{kamland_nn_to_nunubar}
\beq
(\tau/B)_{nn \to {\rm inv.}} > 1.4 \times 10^{30} \ {\rm yr}
\label{tau_nn_to_inv_kamland}
\eeq
per ${}^{12}$C nucleus, where ``inv." denotes an invisible final state, e.g.,
one with two neutral, weakly interacting particles which do not decay in the
detector (and which could be $\nu\nu$, $\nu\bar\nu$, or $\bar\nu\bar\nu$, with
undetermined flavors). Since the final-state (anti)neutrinos were not observed,
the limit (\ref{tau_nn_to_inv_kamland}) applies to all of these
possibilities. For the case where the final state is $\nu_\ell \bar\nu_\ell$,
our estimated lower bound in (\ref{tau_limit_nn_to_nunubar}) is considerably
stronger than the direct experimental limit (\ref{tau_nn_to_inv_kamland}).

\begin{figure}
  \begin{center}
    \includegraphics[height=6cm]{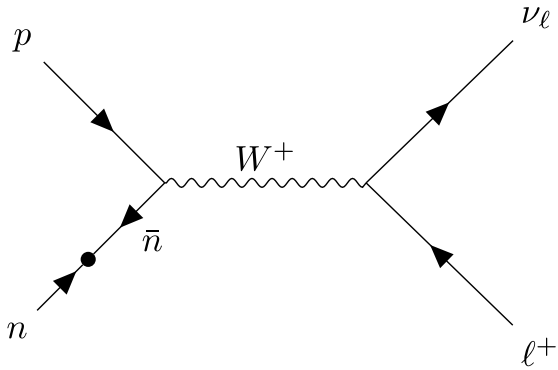}
  \end{center}
\caption{Feynman diagram for $np \to \ell^+ \nu_\ell$, where 
$\ell = e, \ \mu, \ \tau$.} 
\label{dinucleon_to_ellnu_figure}
\end{figure}

Finally, we derive a relation between the rates for $np \to \pi^+\pi^0$ and $np
\to \ell^+ \nu_\ell$, where $\ell^+ = e^+, \ \mu^+, \ \tau^+$.  At tree level,
the amplitude $np \to \ell^+ \nu_\ell$ arises from the process in which the
$|\bar n\rangle$ component in $|n\rangle_{phys.}$ annihilates with a
neighboring proton to produce a virtual $W^+$ boson which then yields the
final-state $\ell^+ \nu_\ell$ pair. This is shown in
Fig. \ref{dinucleon_to_ellnu_figure}.  Denoting the four-momenta of the 
$\nu_\ell$ and $\ell^+$ as $p_2$ and $p_1$, we write 
\beqs
A_{np \to \ell^+ \nu_\ell} &=&
(\delta m) \frac{G_F}{\sqrt{2}} \, \langle 0 | J_W^\lambda|  \bar n p\rangle 
\, [\bar u(p_2)\gamma_\lambda(1-\gamma_5)v(p_1)] \ . \cr\cr
&& 
\label{amp_np_to_ellnu}
\eeqs
The initial $np$ state is a mixture of
$I=0$ and $I=1$ isospin states. The $I=0$ state is analogous to the deuteron,
with $S=1$ and dominantly $L=0$, whence $J=1$. The $I=1$ $np$ state has
dominantly $L=0$, $S=0$, and hence $J=0$, leading to severe helicity
suppression of the decays if $\ell^+=e^+$ or $\ell^+ = \mu^+$, although this
helicity suppression not so severe for $np \to \tau^+ \nu_\tau$. In contrast,
the decays $np \to \ell^+ \nu_\ell$ from the initial $np$ states with $J \ne 0$
are not helicity-suppressed. This is similar to the fact that there is no
helicity suppression in the leptonic decays of a real $W$ boson.  It is thus
expected that the dominant contribution to $np \to \ell^+ \nu_\ell$ arises from
the $I=0$, $J=1$ component of the initial $np$ state. 
We thus estimate 
\beq
\Gamma_{np \to \ell^+ \nu_\ell} \sim
(G_F s)^2 \, \frac{R_2^{(\ell^+\nu_\ell)}}{R_2^{(\pi^+\pi^0)}} \, 
\Gamma_{np \to \pi^+\pi^0}
\label{gamma_np_to_ellnu}
\eeq
The phase space factor for $np \to \ell^+ \nu_\ell$ decay is 
$R_2^{(\ell^+\nu_\ell)} = (8\pi)^{-1} \, [1-m_\ell^2/(2m_N)^2]$. The 
expression in square brackets has the respective values 1.0000, \ 0.9969, \ 
and 0.1047 for $\ell=e, \ \mu, \ \tau$.  In the decay
$np \to \pi^+\pi^0$, $R_2^{(\pi^+\pi^0)} = (8\pi)^{-1}(0.9893)$. 
Combining Eq. (\ref{gamma_np_to_ellnu}) with these values for the phase space
factors and the experimental limit (\ref{tau_np_to_pippi0}), 
we obtain the rough lower bounds, per ${}^{16}$O nucleus, 
\beqs
(\tau/B)_{np \to \ell^+ \nu_\ell}  \gsim 10^{41} \ {\rm yrs} 
\quad {\rm for} \ \ell=e, \ \mu 
\label{tau_limit_np_to_ellnu}
\eeqs
and
\beq
(\tau/B)_{np \to \tau^+ \nu_\tau} \gsim 10^{42} \ {\rm yr} \ . 
\label{tau_limit_np_to_taunu}
\eeq
The SK experiment has reported the limits \cite{sk_np_to_ellnu}
\beq
(\tau/B)_{np \to e^+x } > 2.6 \times 10^{32} \ {\rm yr}
\label{sk_np_to_enu}
\eeq
and
\beq
(\tau/B)_{np \to \mu^+x } > 2.2 \times 10^{32} \ {\rm yr} 
\label{sk_np_to_munu}
\eeq
per ${}^{16}$O nucleus, where $x$ denotes a neutrino or antineutrino (of
undetermined flavor). For the cases in which $x=\nu_e$ in (\ref{sk_np_to_enu})
and $x=\nu_\mu$ in (\ref{sk_np_to_munu}), our bounds are much stronger than
these limits from direct experimental searches.  It was pointed out in
\cite{bryman} that data from existing searches for nucleon and dinucleon 
decays into
multilepton final states involving $e^+$ and $\mu^+$ plus (anti)neutrinos could
be retroactively analyzed to set a limit on the decay $np \to \tau^+
\bar\nu_\tau$, since the $\tau^+$ could decay as $\tau^+ \to \bar\nu_\tau
\ell^+ \nu_\ell$ with $\ell=e$ or $\ell=\mu$.  Ref. \cite{bryman} carried out
such an analysis and obtained a lower bound
$(\tau/B)_{np \to \tau^+\bar\nu_\tau} > 1 \times 10^{30}$ yr per ${}^{16}$O
nucleus.  Subsequently, from a direct search, SK obtained the limit
\cite{sk_np_to_ellnu}
\beq
(\tau/B)_{np \to \tau^+ x} > 2.9 \times 10^{31} \ {\rm yr} 
\label{sk_np_to_taunu}
\eeq
per ${}^{16}$O nucleus, where $x$ is a neutrino or antineutrino (of
undetermined flavor).  For the case in which $x=\nu_\tau$, our bound
(\ref{tau_limit_np_to_taunu}) is much stronger than this direct limit.  As is
evident from our derivations, our limits constrain dinucleon decays that have
$\Delta L = 0$. They do not constrain dinucleon decays with $\Delta L \ne 0$,
such as the $\Delta L=-2$ decays $nn \to \bar\nu_\ell \bar\nu_{\ell'}$ and $np
\to \tau^+\bar\nu_\tau$ or the $\Delta L = +2$ decay $nn \to \nu_\ell
\nu_\ell$. Using similar methods, we have derived improved upper bounds on
several decay models of individual protons and bound neutrons.  These are
reported elsewhere.

This research was supported in part by the NSF Grants NSF-PHY-1620628 and 
NSF-PHY-1915093 (R.S.). 

% ======================================================================

% ======================================================================


\begin{thebibliography}{99}

% 1
\bibitem{sakharov} 
A. D. Sakharov, JETP Lett. B {\bf 91} (1967) 24  
[Zh. Eksp. Teor. Fiz. Pis'ma {\bf 5} (1967) 32]. 

% 2 
\bibitem{kuzmin}
V. Kuzmin, JETP Lett. {\bf 12} (1970) 228 
[Zh. Eksp. Theor. Fiz. Pis'ma {\bf 12} (1970) 335]. 

% 3
\bibitem{ennb}
R. Mohapatra and R. Marshak, Phys. Rev. Lett. {\bf 44} (1980) 1316; 
L.-N. Chang and N.-P. Chang, Phys. Lett. {\bf B92} (1980) 103;
Phys. Rev. Lett. {\bf 45} (1980) 1540;
T. K. Kuo and S. Love, Phys. Rev. Lett. {\bf 45} (1980) 93; 
R. Cowsik and S. Nussinov, Phys. Lett. {\bf B101} (1981) 237. 

% 4 
\bibitem{nnb82}
S. Rao and R. Shrock, Phys. Lett. {\bf B116} (1982) 238. 

% 5 
\bibitem{nnb84}
S. Rao and R. Shrock, Nucl. Phys. {\bf B232} (1984) 143.

% 6
\bibitem{ill}
M. Baldo-Ceolin et al., Zeit. f. Phys. C {\bf 63} (1994) 409. 

% 7
\bibitem{nnb02}
S. Nussinov and R. Shrock, Phys. Rev. Lett. {\bf 88} (2002) 171601. 

% 8
\bibitem{soudan_nnbar}
J. Chung et al. (Soudan Collab.), Phys. Rev. D {\bf 66} (2002) 032004. 

% 9
\bibitem{friedman_gal}
E. Friedman and A. Gal, Phys. Rev. D {\bf 78} (2008) 016002. 

% 10
\bibitem{wise}
J. M. Arnold, B. Fornal, and M. B. Wise, Phys. Rev. D {\bf 87} (2013) 075004.

% 11 nnbar-whitepaper
\bibitem{nnbar_white}
D. C. Phillips et al., Phys. Repts., {\bf 612} (2015) 1. 

% 12 frejus
\bibitem{frejus91}
C. Berger et al. (Fr\'ejus Collab.), Phys. Lett. B {\bf 269} (1991) 227. 

% 13
\bibitem{takita86}
M. Takita et al. (Kamiokande Collab.), Phys. Rev. D {\bf 34}, 902 (1986). 

% 14
\bibitem{chung02}
J. Chung et al. (Soudan Collab.), Phys. Rev. D {\bf 66}, 032004
(2002) (iron).

% 15
\bibitem{aharmim17}
B. Aharmim et al. (SNO Collab.) Phys. Rev. D {\bf 96}, 092005 (2017).

% 16
\bibitem{sk_nnbar}
K. Abe et al. (SuperKamiokande Collab.), Phys. Rev. D {\bf 91} (2015) 072006.

% 17 sk dinucleon to pions 
\bibitem{sk_dinucleon_to_pions}
J. Gustafson et al. (SuperKamiokande Collab.), Phys. Rev. D {\bf 91}
(2015) 072009. 

% 18  two-nucleon (B-L)-conserving reactions involving tau leptons 
\bibitem{bryman}
D. Bryman, Phys. Lett. B {\bf 733} (2014) 190. 

% 19 SuperK np to tau X limit 
\bibitem{sk_np_to_ellnu} 
V. Takhistov et al. (SuperKamiokande Collab.), Phys. Rev. Lett. {\bf 115} 
(2015) 121803. 

% 20
\bibitem{sk_dinucleon_to_ellell}
S. Sussman et al. (SuperKamiokande Collab.), arXiv:1811.12430. 

% 21
\bibitem{kamland_nn_to_nunubar}
T. Araki et al. (KamLAND Collab.), Phys. Rev. Lett. {\bf 96} (2006) 101802.  

% 22
\bibitem{sno_invisible}
M. Anderson et al. (SNO+ Collab.), Phys. Rev. D {\bf 99} (2019) 032008. 

\end{thebibliography}
\end{document}